%% file: main.tex
\documentclass[reprint,amsmath,amssymb,aps, superscriptaddress]{revtex4-2}

\usepackage{graphicx}
\usepackage[caption=false]{subfig}
\usepackage{float}
\usepackage{xcolor}
\usepackage{booktabs}
\usepackage{hyperref}
\usepackage[capitalize]{cleveref}

\hypersetup{
    colorlinks,
    linkcolor={blue!50!black},
    citecolor={blue!50!black},
    urlcolor={blue!80!black}
}

\begin{document}
\title{Machine Learning the Entropy to Estimate Free Energy Differences without Sampling Transitions}
\date{November 2025}
\begin{abstract}
Thermodynamic phase transitions, a central concept in physics and chemistry, are typically controlled by an interplay of enthalpic and entropic contributions. In most cases, the estimation of the enthalpy in simulations is straightforward but evaluating the entropy is notoriously hard. 
As a result, it is common to induce transitions between the metastable states and estimate their relative occupancies, from which the free energy difference can be inferred. However, for systems with large free energy barriers, sampling these transitions is a significant computational challenge. Dedicated enhanced sampling algorithms require significant prior knowledge of the slow modes governing the transition, which is typically unavailable. 
We present an alternative approach, which only uses short simulations of each phase separately. We achieve this by employing a recently developed deep learning model for estimating the entropy and hence the free energy of each metastable state.  
We benchmark our approach calculating the free energies of crystalline and liquid metals. Our method features state-of-the-art precision in estimating the melting transition temperature in Na and Al without requiring any prior information or simulation of the transition pathway itself.
\end{abstract}

 \author{Yamin Ben-Shimon}
 \affiliation{School of Physics, Tel Aviv University, Tel Aviv 6997801, Israel.}
 \affiliation{The Center for Physics and Chemistry of Living Systems, Tel Aviv University, Tel Aviv 6997801, Israel.}

 \author{Barak Hirshberg}
 \email{hirshb@tauex.tau.ac.il}
\affiliation{School of Chemistry, Tel Aviv University, Tel Aviv 6997801, Israel.}
\affiliation{Center for Computational Molecular and Materials Science, Tel Aviv University, Tel Aviv 6997801, Israel.}
\affiliation{The Center for Physics and Chemistry of Living Systems, Tel Aviv University, Tel Aviv 6997801, Israel.}

 \author{Yohai Bar-Sinai}
 \email{ybarsinai@gmail.com}
  \affiliation{School of Physics, Tel Aviv University, Tel Aviv 6997801, Israel.}
  \affiliation{The Center for Physics and Chemistry of Living Systems, Tel Aviv University, Tel Aviv 6997801, Israel.}
  
\maketitle

\section{Introduction}
Estimating the relative stability of different phases that are separated by high free energy barriers is a challenging task. To compare between phases, simulations must ergodically sample the relevant regions of phase space. However, due to the high free energy barriers, transitions are rare and metastable states persist for long timescales. Consequently, standard simulation techniques cannot sample transitions within feasible computational times~\cite{frenkel2023understanding}.

Various approaches were developed to enhance the sampling efficiency of rare transitions. Many of them, such as Metadynamics~\cite{barducci2008well,barducci2011metadynamics,valsson2016enhancing,sutto2012new,bussi2020using}, Umbrella Sampling~\cite{kastner2011umbrella,torrie1977nonphysical}, Gaussian accelerated molecular dynamics~\cite{miao2015gaussian}, or on-the-fly probability enhanced sampling~\cite{invernizzi2020rethinking,invernizzi2020unified,invernizzi2022exploration,Invernizzi2021}, bias the system to induce transitions. These methods rely heavily on the identification of suitable collective variables (CVs), which should represent the slow modes of the system. This limits their applicability, since identifying appropriate CVs in condensed phases is challenging~\cite{rogal2019neural}.

If, instead, we could estimate the entropy of each phase directly, without inducing transitions between the phases, we would be able to evaluate free energy differences using the thermodynamic relation 
\begin{equation}
\Delta g = \Delta h - T \Delta s \ .
\label{eq:DeltaFrelation}
\end{equation}
In~\cref{eq:DeltaFrelation}, $\Delta g$ is the free energy density difference, $\Delta h$ is the enthalpy density difference, and $\Delta s$ is the entropy density difference between the phases~\cite{callen1980thermodynamics}. Note that this procedure is agnostic to the details of the transition path between metastable states. The challenge with this approach is that while calculating the enthalpy is straightforward, evaluating the entropy is generally an open problem: The enthalpy depends only on the energy, pressure and volume, which are routinely evaluated in simulations, but computing the entropy requires the partition function.

A comprehensive review of methods for direct entropy estimation is beyond the scope of this paper, and we focus on select recent works. Avinery et al.~\cite{avinery2019universal} and Martiniani et al.~\cite{martiniani2019quantifying} leveraged established lossless compression algorithms to bound the information content of a sequence of sampled microstates, which is equivalent to the thermodynamic entropy. The applicability of these methods depends on the performance of the underlying compression algorithm to the problem at hand~\cite{liu2024deltagzip}.
Since compression algorithms treat data as a one-dimensional string, this approach works well for systems with a natural sequential structure, but face difficulties  when forcing more complicated geometries into a 1D sequence~\cite{Zu2020}.
More recently, Sorkin et al.~developed a novel upper bound on the entropy, based on correlations between various degrees of freedom~\cite{sorkin2023detecting,sorkin2023resolving}.
These methods are promising, and successfully estimated the entropy in model systems, but were not applied to molecular simulations of phases separated by high free energy barriers.

Machine learning algorithms have also been successfully used to estimate the entropy in physical systems. 
Gelman et al.~estimated the probability density of microstates directly by training an auto-regressive model~\cite{gelman2024nonequilibrium}. Then, they calculated the entropy from the log-probability. However, this method is limited to lattice systems in two dimensions, and does not scale well with system size.
Generative models have recently been proposed as a method to directly sample from the Boltzmann distribution~\cite{noe2019boltzmann}, where free energy can be inferred from the relative frequencies of the metastable states. Covino et al. even used diffusion models to generate whole trajectories recently~\cite{petersen2023dynamicsdiffusion}. While promising, they have so far been applied mostly to small molecules~\cite{klein2024transferable}.
Most relevant to this work is a method called MICE by Nir et al.~\cite{Nir_2020}. They estimated the entropy by mapping the problem to an iterative process of mutual information (MI) estimation at different length scales. The latter was done by representing the MI as an optimization problem, parameterized by a neural network, as proposed by Belghazi et al.~\cite{belghazi2021minemutualinformationneural}.

We adopt the MICE approach and combine it with molecular dynamics (MD) simulations for the first time. This offers a new method to estimate the entropy and relative stability of phases separated by high free-energy barriers. Our computationally efficient approach for learning the entropy requires only short simulations of each phase separately. It offers an accurate calculation of the critical temperature of melting phase transitions without the need for prior knowledge of the system (CVs) and without sampling transitions. We demonstrate the usefulness of our approach in estimating the melting temperature of Na. Without any changes to the model architecture or hyperparameters, our approach also accurately estimates the melting temperature of Al.

\section{Methodology}
The central concept underlying MICE is that the entropy can be estimated as the sum of MI contributions at different length scales. Consider two random variables $A$ and $B$, their mutual information $MI(A,B)$ is defined as:  
\begin{equation}\label{eq:entropy_def_S}
    S(A,B) = S(A) + S(B) - MI(A,B)\ ,
\end{equation}
where $S(A,B)$ is the entropy of the joint distribution and $S(A), S(B)$ are the marginal entropies.
MICE uses this relationship to replace the estimation of the total entropy with that of the entropies of the subsystems and their MI. 
This approach is useful because the computational complexity of entropy estimation is exponential in the system size.
By breaking the system into smaller, more manageable parts, while keeping track of the MI between them, the overall computational demands are significantly reduced.

Consider a physical system $X_0$ of volume $V_0$, which is split into two equal parts of volume $V_1=\frac{1}{2}V_0$. We treat $X_0$ as a random variable drawn from the Boltzmann distribution.
If the system is translationally invariant (which is the case far from boundaries or with periodic boundary conditions), the two halves are statistically identical and can be both denoted as $X_1$. With this setup, \cref{eq:entropy_def_S} simplifies to:
\begin{equation}\label{eq:mice_def_S}
    S(X_0) = 2S(X_1)-MI(X_1) \ ,
\end{equation}
where $MI(X_1)$ is the mutual information between the two subsystems $X_1$. 
This process can be repeated for increasingly smaller systems, resulting in
\begin{equation}\label{eq:mice_sum_iterations}
    s(X_0) = s(X_m) - \frac{1}{2}\sum_{k=1}^{m} \frac{MI(X_k)}{V_k} \ ,
\end{equation}
where $s(X_k)=S(X_k)/V_k$ is the entropy density for a subsystem $X_k$ with volume $V_k$, see~\cref{sec:mice_derivation} for details and derivation. For later use, we also denote the interface area between two neighboring $X_k$ systems by $A_k$.
\cref{eq:mice_sum_iterations} decomposes the entropy $S$ into contributions from different length scales. For the smallest subdivision, $s(X_m)$ can be calculated directly, either by brute-force enumeration or other methods. In our application, we continue the division until $X_m$ is small enough such that it contains a single particle on average. Since the volume of $X_k$ decreases exponentially with $k$, the required number of subdivisions is logarithmic in the system size.

\begin{figure*}[t]
\centering
\includegraphics[width=1
\linewidth]{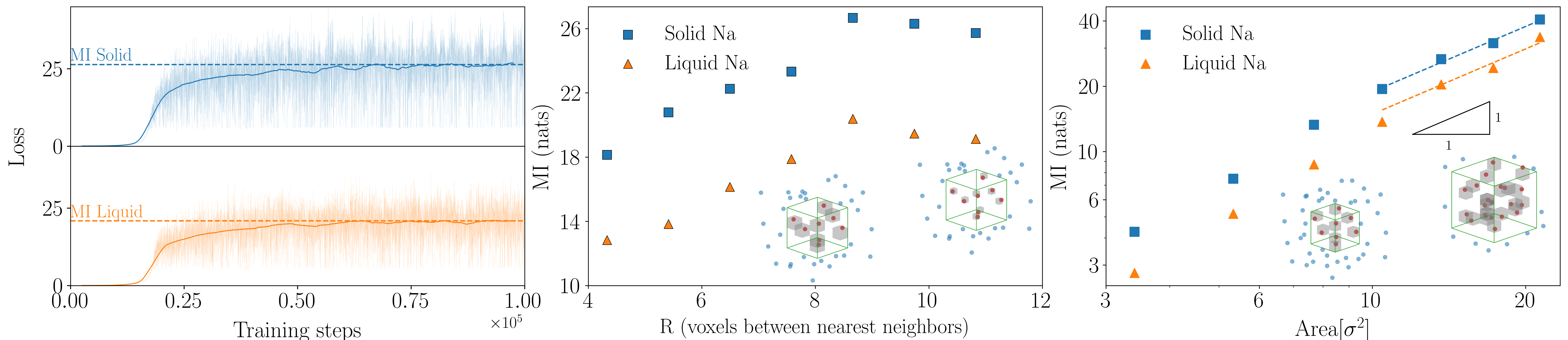}
\caption{\textbf{MI estimations for liquid and crystalline Na} (left) Training curves for solid (blue, upper plot) and liquid (orange, bottom plot) Na close to $T_m$. Bold lines represent running averages over $5000$ steps and dashed line are the converged values over last $10^4$ steps. (middle) MI estimation for a system of fixed physical size as a function of the spatial resolution. The estimate plateaus at high resolutions. (right) Estimation of MI between two halves of a cubic subsystem, as a function of the interface area. The dashed lines show a linear trendline, showing that large systems obey an area law. The corresponding figures for Al data are presented in~\cref{sec:aluminum}.}
\label{fig:res_Na}
\end{figure*}

To estimate $MI(X_k)$, we employ the Mutual Information Neural Estimator (MINE) \cite{belghazi2021minemutualinformationneural}, which we describe here briefly. The method relies on the representation of $MI$ between two random variables $X$ and $Y$ as the supremum over all real functions $\mathcal{T}(X,Y)$~\cite{donsker1975asymptotic},
\begin{equation}
\begin{split}
MI(X,Y)&=\\\sup _{\mathcal{T}}\Big\{ \mathbb{E}_{\mathbb{P}_{X Y}}&\left[\mathcal{T}(X,Y)\right]-\log \left(\mathbb{E}_{\mathbb{P}_X \otimes \mathbb{P}_Y}\left[e^{\mathcal{T}(X,Y)}\right]\right)\Big\} \ .
\end{split}
\label{eq:mine}
\end{equation} 
Here, $\mathbb{P}_{XY}$ is the joint probability distribution of the two variables and $\mathbb{P}_X \otimes \mathbb{P}_Y$ is the product of their marginal distributions. 
Then, $MI$ is estimated by parameterizing $\mathcal{T}$ with a neural network, and optimizing its weights. Since there is no guarantee that the optimization will find the global supremum, this procedure bounds the true MI from below, resulting in an upper bound on the entropy (~\cref{eq:mice_sum_iterations}).

To apply MICE, we first sample equilibrium snapshots for each phase by running separate, standard MD simulations. 
Our $X_0$ is an MD simulation box with periodic boundary conditions, see below for details.
We obtain samples for each subdivision $k$ from $\mathbb{P}_{X_k X_k}$ by choosing a volume $V_k$ in a random location out of the original simulation box. Samples from $\mathbb{P}_{X_k} \otimes \mathbb{P}_{X_k}$ are generated by stitching together two halves of randomly chosen samples from the joint distribution. 
We then train a neural network to optimize \cref{eq:mine}. 
\cref{fig:res_Na}a shows representative training curves for a typical subdivision of solid and liquid Na. While the training curves are noisy~\cite{choi2022combating}, a running average shows convergence to different $MI$ estimations for the two phases, which we take as the estimate for $MI(X_k)$.
We repeat this procedure for $k=1,...,m$ and for each phase separately. The entropy of $X_0$ is obtained by summing over all subsystems through~\cref{eq:mice_sum_iterations}. 

A key component of the method is the choice of representation of the configurations $X_k$ for training. Many choices are possible, and here we adopt the representation used in~\cite{Nir_2020}: a cubic grid with $n^3$ voxels, each of side length $p$. The value of a voxel is 1 if it contains at least one atom, and zero otherwise.
As $n$ increases, the embedding resolution increases, which we measure through the dimensionless number $R = \sigma/p$, where $\sigma$ is the nearest-neighbor distance in the experimental crystal structure. \cref{fig:res_Na}b shows that the $MI$ converges for $R\approx10$ for Na. 

Another important point is that, since we are dealing with bulk properties, $X_0$  of \cref{eq:mice_sum_iterations} should be large enough to represent an effectively infinite system.
The size of $X_0$ should thus be chosen such that it minimizes finite-size effects on the one hand, but is numerically manageable on the other hand.
This ``sweet spot'' can be obtained by using the asymptotic area law of the entropy~\cite{wolf2008area}: At length scales larger than the longest correlation length, $MI(X_k)$ should scale linearly with $A_k$.
Indeed, \cref{fig:res_Na}c shows $MI$ as a function of $A_k$ for increasing system size. We find that for small systems $MI$ scales super-linearly with $A_k$, and becomes linear for larger systems, see dashed lines in \cref{fig:res_Na}c. 
We choose $X_0$ of \cref{eq:mice_sum_iterations} to be slightly larger than this crossover size.
This allows us to take the limit of $X_0 \to \infty$ analytically, under the assumption that for systems larger than $X_0$, we have $MI(X_k)\propto A_k$. As shown in the SI, this procedure results in 
\begin{equation}\label{eq:mice_extrapolation}
    s \;=\; s(X_0) \;-\; 3\,\frac{MI(X_{0})}{V_{0}}\ .
\end{equation}

\section{Computational Details}
To benchmark our method, we performed well-tempered Metadynamics (WT-MetaD) simulations following the protocol of Piaggi et al.~\cite{piaggi2017enhancing} using LAMMPS (15Jun2023)~\cite{thompson2022lammps} and PLUMED 2.8.3~\cite{bonomi2009plumed, plumed2019promoting, tribello2014plumed}. We ran NPT WT-MetaD simulations in the temperature ranges 300–400 K for Na and 850–950 K for Al. We used the pair entropy $s_{2}$ and the enthalpy per atom as CVs. 
For Na, we deposited Gaussian hills with an initial height of 2.5 kJ mol$^{-1}$ every 500 steps with widths of 0.2 kJ mol$^{-1}$ (enthalpy per atom) and 0.1 ($s_{2}$). For Al, we used a larger initial height of 7.5 kJ mol$^{-1}$ and widths of 0.3 kJ mol$^{-1}$ (enthalpy) and 0.1 ($s_{2}$), keeping the same deposition stride. A bias factor of 30 was used for both systems. Convergence was assessed as in Ref.~\cite{piaggi2017enhancing}. 
We used a Parrinello-Rahman barostat at 1~bar acting on a triclinic simulation cell, allowing isotropic volume fluctuations while relaxing shear components to zero. Simulations were performed on a 5x5x5 supercell for Na with 250 atoms and a 4x4x4 supercell for Al with 256 atoms.

To generate data for MICE, we performed unbiased MD simulations of Na and Al separately in the solid (bcc and fcc, respectively) and liquid phases with LAMMPS. Collective variables were tracked using PLUMED 2.8.3~\cite{bonomi2009plumed, plumed2019promoting, tribello2014plumed}, but not used for biasing the dynamics. Temperature was maintained close to the experimental melting points, 350\,K for Na and 900\,K for Al using a stochastic velocity–rescaling thermostat~\cite{bussi2007canonical} with a relaxation time of 0.1 ps. 
The pressure was set to 1~bar using the isotropic Parrinello–Rahman barostat on an orthorhombic cell~\cite{parrinello1981polymorphic} with a relaxation time of 10 ps. Simulations were performed on an 8x8x8 supercell with 1024 atoms for Na and 2048 atoms for Al. 
We ran 20 independent replicas with different seeds for each element and phase, and sampled configurations every 5 ps, for a total of 40000 data points, which were evenly split into training and validation sets. 

Training data was obtained by taking sub-volumes of the simulation box. $X_0$, the biggest subsystem used in \cref{eq:mice_sum_iterations}, was taken to be a cubic volume containing, on average, 60 atoms for Na and 50 for Al, i.e.~about a third of the linear size of the simulation box, to avoid periodic imaging effects.
For each snapshot, a sub-volume was chosen at a random location and orientation, uniform over the simulation box and SO(3), respectively. 
Each such sub-volume was discretized into a binary $n \times n \times n$ voxel lattice, marking cells 1 if they contain a particle and 0 otherwise. The resolution was chosen to be fine enough such that a voxel never contains more than one particle. Samples of smaller volumes, $X_1, X_2, $ etc, were obtained by cropping samples of $X_0$. 

The function $\mathcal{T}$ was parameterized with a 3D convolutional network composed of four successive convolutional blocks with LeakyReLU activation, 0.15 convolutional dropout and adaptive max pooling to target output sizes $(20,10,5,2)$. 
Starting with $c=22$ channels, the width is scaled by a 2.5 factor at each block. Finally three fully connected layers with 0.3 dropout \cite{srivastava14Dropout} were used. 
Weights were initialized with Xavier uniform~\cite{glorot2010understanding}. 
For optimization, we used Adam (learning rate $3\times10^{-5}$, batch size~1200) for $1.5\times10^{5}$ batches. 
An exponential moving average estimator was maintained with rate $2.5\times10^{-7}$ and starting value $10^{-5}$. 

We note our experiments showed that, at least for smaller subsystems, equivalent performance in MI estimation can be obtained with smaller networks. However, since the goal of this work is just to demonstrate the applicability of the method, we wanted to avoid any possible effects of hyperparameter tuning, and chose to use the same architecture for all subsystem sizes. 

\begin{figure}[t]
\centering
\includegraphics[width=1
\linewidth]{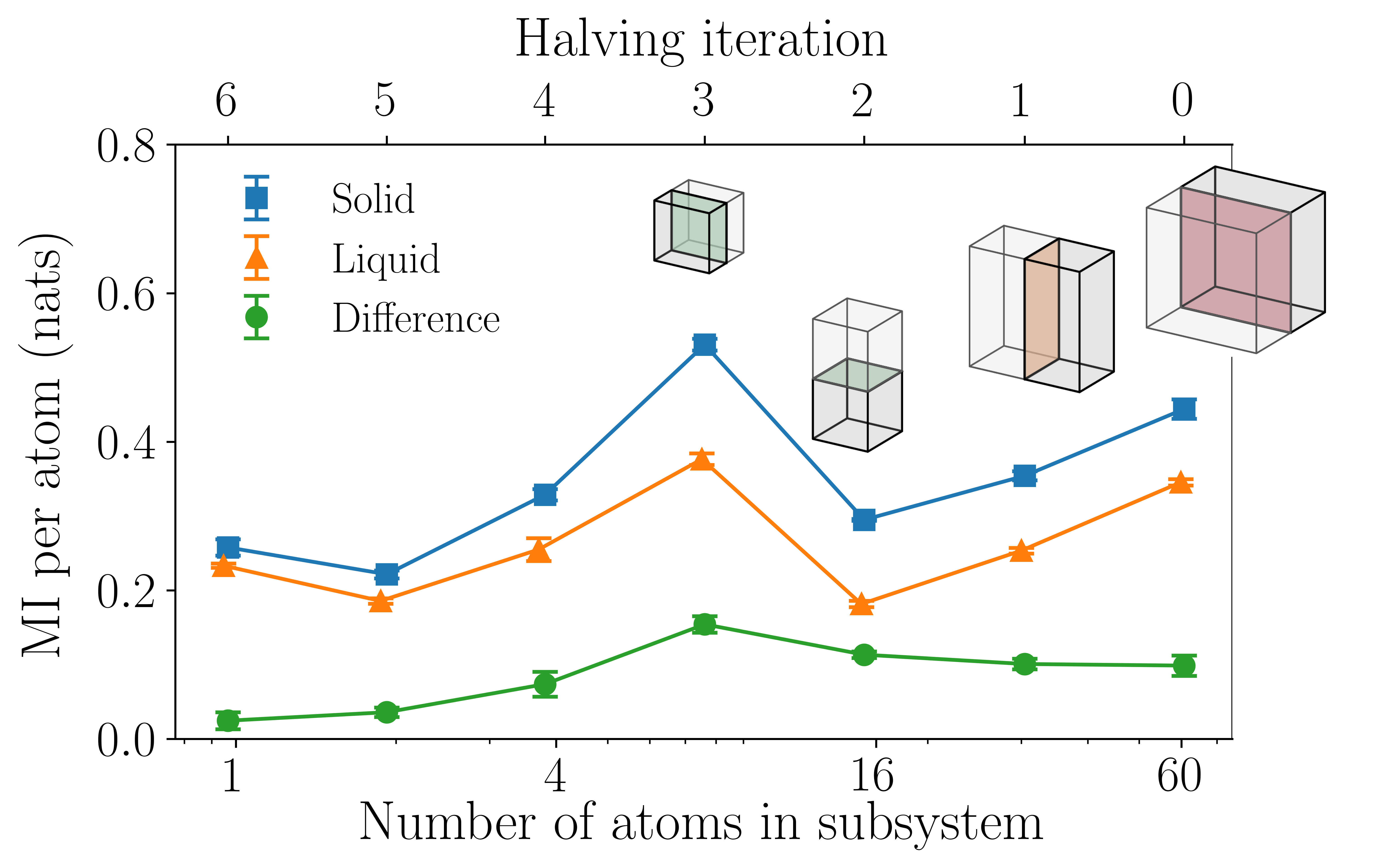}
\caption{MI density across successive partitions. Starting from the rightmost panel, the system is halved along a different dimension at each step, and MI is computed across the resulting interface. Because the three cut directions cycle, the interface area is unchanged every third partition while the subsystem volume halves at every step (e.g., iterations 2-3). Consequently, the MI density doubles between such iterations.
}
\label{fig:halving}
\end{figure}
\section{Results and Discussion}

\begin{figure}[t]
\centering
\includegraphics[width=1.
\linewidth]{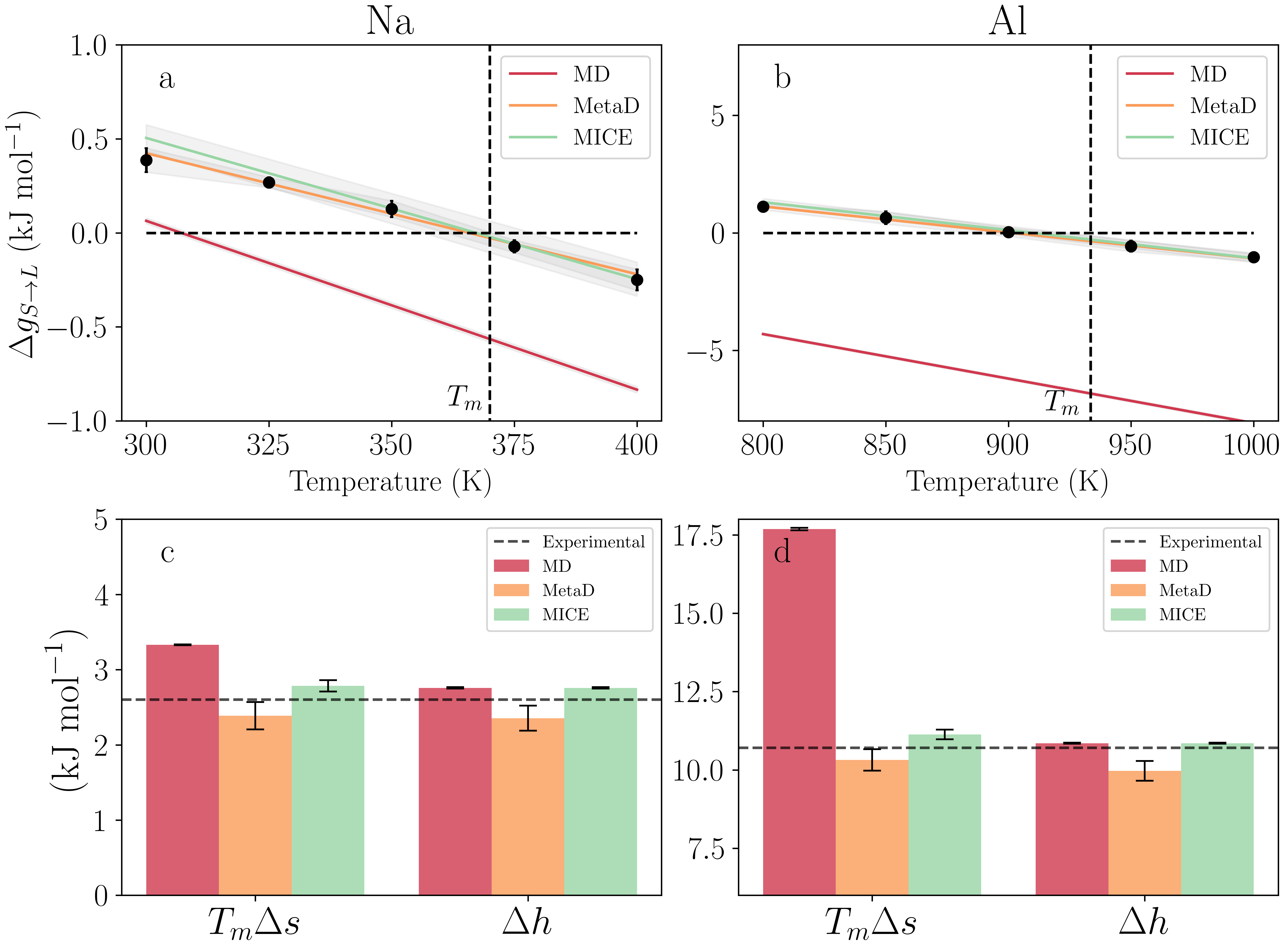}
\caption{(a-b) Gibbs free energy of melting vs temperature. The data for classical MD and MICE are generated from sampled $\Delta h$ and the entropy estimates $\Delta s_2$ and $\Delta s_{MICE}$ correspondingly. WT-MetaD data is shown by the black points, and the linear fit through them produces estimates of $\Delta h$ and $\Delta s$. (c-d) $\Delta h$ and $T_m\Delta s$ calculated from classical MD, WT-MetaD and MICE (enthalpy estimations for MICE and MD are identical by construction).}
\label{fig:dg}
\end{figure}

\cref{fig:halving} shows the MI per atom for solid and liquid Na for the different subsystems divisions of the MICE procedure (~\cref{eq:mice_sum_iterations}). As expected, since structural information is encoded in long wavelengths, the difference in MI between the solid and liquid phases decreases for smaller subsystems with fewer atoms and vanishes for subsystems containing only one atom. The jump in MI per atom every three divisions corresponds to a divide with a large aspect ratio of interface area to volume (~\cref{fig:halving} (inset)).

With~\cref{eq:mice_sum_iterations} and~\cref{eq:mice_extrapolation}, these MI estimations are used to calculate the specific entropy density difference between the phases, $\Delta s$. Results are shown in \cref{fig:dg}~a,c (labeled ``MICE"), and compared to the estimates calculated from the same unbiased single-phase simulations using $s_2$ (labeled ``MD"). In addition, we show the WT-MetaD estimates which are obtained by fitting a linear trend to $\Delta g$ values at different temperatures and using \cref{eq:DeltaFrelation} (labeled ``MetaD"). Enthalpy estimation in all three methods is also plotted, and is identical for MD and MICE simulations by definition, since they both use time averages of the same unbiased dynamics. 

Unsurprisingly, it is seen that $s_2$ provides a poor estimation of the entropy difference. If combined with the measured enthalpy difference, it would predict a melting temperature of about 300K, a relative error of $\sim 25\%$. 
WT-MetaD simulations underestimate the entropy, but the resulting prediction of $T_m$ lies within several kelvin of the experimental value, due to a cancellation of a similar error in $\Delta h$. This tendency was reported for crystallization transitions and is consistent with the known bias compression factor of WT-MetaD~\cite{barducci2011metadynamics,piaggi2017enhancing}. 
MICE, on the other hand, produces a modest overestimation in $\Delta s$, and an unbiased estimation of $\Delta h$, which yield a melting temperature closer to the experimental value. 

After establishing that our method works reliably for the melting transition of Na, we applied exactly the same network, with identical hyper-parameters settings (layer sizes, dropout, etc.), to a similar dataset for solid and liquid Al. This data set was prepared with a similar voxel resolution and mean atomic density. Without any additional tuning or architectural changes, the model converged smoothly and delivered MI estimates, and consequently $\Delta s$ estimates, which were used to predict the melting temperature, mirroring the accuracy obtained for Na, as shown in ~\cref{fig:dg} b,d.
This portability emphasizes the robustness and generality of our representation and training protocol.
Here too, $\Delta s_2$ is a poor approximation of the entropy, and WT-MetaD predicts a relatively precise melting temperature due to a systematic bias for both the enthalpy and entropy difference. 

To conclude, we present a machine-learning approach to estimate free-energy differences between metastable states that are separated by high barrier without sampling transitions between them. It does so by formulating the entropy difference as a sum over MI at different lengths scales. Then, the contribution of each length scale is obtained by optimizing a convolutional neural network. Our approach predicts the melting temperature of Na and Al using only short simulations of each phase separately, and without requiring collective variables or samples of the transitions pathway.

While MICE shows state-of-the-art accuracy in entropy estimation for atomic systems, there is still much room for improvement. First, the voxel-based encoding may introduce spurious correlations and is not natural for molecular data: it neglects symmetries and local bonding, potentially distorting or omitting relevant information. This could be improved by adopting a graph-like representation with molecularly informed embeddings that better respect symmetry and local structure, as was done in a different context~\cite{batzner20223,satorras2021n,gasteiger2020directional}.
Second, the MINE estimator at the core of our approach \cite{belghazi2021minemutualinformationneural} is optimized with stochastic, finite-batch gradients, which are biased and can inflate MI in the high-MI regime (~\cref{sec:mine_bias}). Using recently developed MINE variants that address bias may improve performance. These directions will be pursued in future research. 

\section{Acknowledgments}
    Yohai Bar-Sinai is supported by Israel Science Foundation grant No.~1907/22 and by Google Gift grant.
    Barak Hirshberg is supported by the Israel Science Foundation (grants No.~1037/22 and 1312/22) and the Pazy Foundation of the IAEC-UPBC (grant No.~415-2023). 
    Yamin Ben-Shimon is supported by the Tel Aviv University Center for AI and Data Science (TAD).
    We thank Pablo Piaggi for sharing his code and useful discussions.
\clearpage

\bibliographystyle{apsrev4-2}
\bibliography{references}
\input{SI}

\end{document}

%% file: SI.tex
\newpage
\appendix
\section{Extrapolating MICE to bulk properties using the area law} \label{sec:mice_derivation}
We present here more details about the calculation in the MICE algorithm.
First we normalize \cref{eq:mice_def_S}, in order to work with the specific entropy, $s_m = S(X_m)/V_m$, where $V_m$ is the volume of the subsystem $X_m$,

\begin{equation*} \label{eq:mice_def_S_norm}
    s(X_0) = \frac{S(X_0)}{V_0} = \frac{2S(X_1)}{V_0} - \frac{MI(X_1)}{V_0} = s(X_1) -\frac{MI(X_1)}{2V_1}
\end{equation*}
where $V_k = 2^{-k}V_0$ is the volume of $X_k$.
This can be iterated arbitrarily many times, so that after $m$ iterations one obtains
\begin{equation}\label{eq:mice_sum_iterations2}
    s(X_0) = s(X_m) - \frac{1}{2}\sum_{k=1}^{m} \frac{MI(X_k)}{V_k} \ .
\end{equation}
This procedure should be repeated until the entropy of $X_m$ can be computed directly, or is physically uninteresting.

The algorithm described above gives an estimation of the specific entropy of a finite system $X_0$. To deal with bulk properties, one needs to perform a similar calculation, but for successively bigger systems. To this end, we use negative indices to indicate successive expansions and denote

\begin{equation}
\begin{split}
s(X_{-1}) &= s(X_{0}) - \frac{MI(X_{0})}{2V_{0}} \\[6pt]
s(X_{-2}) &= s(X_{-1}) - \frac{MI(X_{-1})}{2V_{-1}}\\
          &= s(X_{0}) - \frac{MI(X_{0})}{2V_{0}} - \frac{MI(X_{-1})}{4V_{0}} \\[6pt]
&\vdots \\[6pt]
s(X_{-m'}) &= s(X_{0}) - \frac{1}{2V_{0}}\sum_{k=0}^{m'-1} 2^{-k}\,MI(X_{-k})
\end{split}
\label{eq:mice_extrapolation_si}
\end{equation}

In contrast to the division procedure, the expansion process has no cut-off, since all system sizes contribute to the entropy density estimate. However, for systems much larger than the correlation length, we can assume that the area law holds, $MI_k = \alpha A_k$, where $A_k$ is the interface area between two neighboring $X_k$ system, and $\alpha$ is a proportionality factor that can be measured directly from \cref{fig:res_Na}c.

This assumption allows the evaluation of the summation in~\cref{eq:mice_extrapolation_si} as $m \to \infty$. 
Note that every third expansion leaves the MI unchanged because the division occurs along an axis perpendicular to the interface (see inset of \cref{fig:halving}), keeping its area constant. Thus, the interfacial areas obey $$\frac{A_{-k}}{A_{0}} = 2^{\,k - \left\lfloor \tfrac{k}{3} \right\rfloor}\ .$$
Using this, and the fact that
$$\lim_{m \to \infty} \sum_{k=0}^{m} 2^{-\left\lfloor \tfrac{k}{3} \right\rfloor} = 6\ ,$$
the thermodynamic limit $m'\to-\infty$ is obtained:
\begin{equation}
    s_{m \to \infty} \;=\; s(X_{0}) \;-\; 3\,\frac{MI(X_{0})}{V_{0}}.
\end{equation}

Combining with \cref{eq:mice_sum_iterations}, we finally obtain the the bulk entropy density 
\begin{equation}
    s \;=\; s(X_m) - \frac{1}{2}\sum_{k=1}^{m} \frac{MI(X_k)}{V_k} \;-\; 3\,\frac{MI(X_{0})}{V_{0}}.
\end{equation}

\section{MINE's bias} \label{sec:mine_bias}
When training with mini-batches, the gradient estimation in MINE training is biased. This is because the gradient of the second term in \cref{eq:mine} is approximated for each batch as
\begin{equation}
\frac{\mathbb{E}_B\!\big[\nabla_\theta \mathcal{T}_\theta \, e^{\mathcal{T}_\theta}\big]}{\mathbb{E}_B\!\big[e^{\mathcal{T}_\theta}\big]}   
\end{equation}
where $\mathbb{E}_B$ denotes the empirical average over a batch~\cite{belghazi2021minemutualinformationneural}. The denominator of this expression is a biased estimator, and for small values of the exponent this can create numerical stability issues.

Belghazi et al.~mitigate the issue by replacing the denominator with an exponential moving average (EMA) of past mini-batch estimates, which reduces variance but leaves a residual bias in the gradient direction \cite{belghazi2021minemutualinformationneural}. More recently, Choi et al.~showed that the EMA fails to correct a drift of the MINE network and exploding exponentials that can skew the MI estimate.  
They introduce a regularization term that fixes the marginal moment, yielding a new family of bounds that suppress drift and variance without relying on an EMA \cite{choi2022combating}.  

In this work we use the original EMA bias reduction and addressed the instability through a simpler but bias-free approach. We choose a large batch size so that the mini-batch denominator is less noisy. 
Although computationally heavier, this approach does not introduce additional bias and keeps MI estimates reliable for both the largest and the smallest subsystems.  

\section{Aluminum} \label{sec:aluminum}
\begin{figure}[H]
\centering
\includegraphics[width=1
\linewidth]{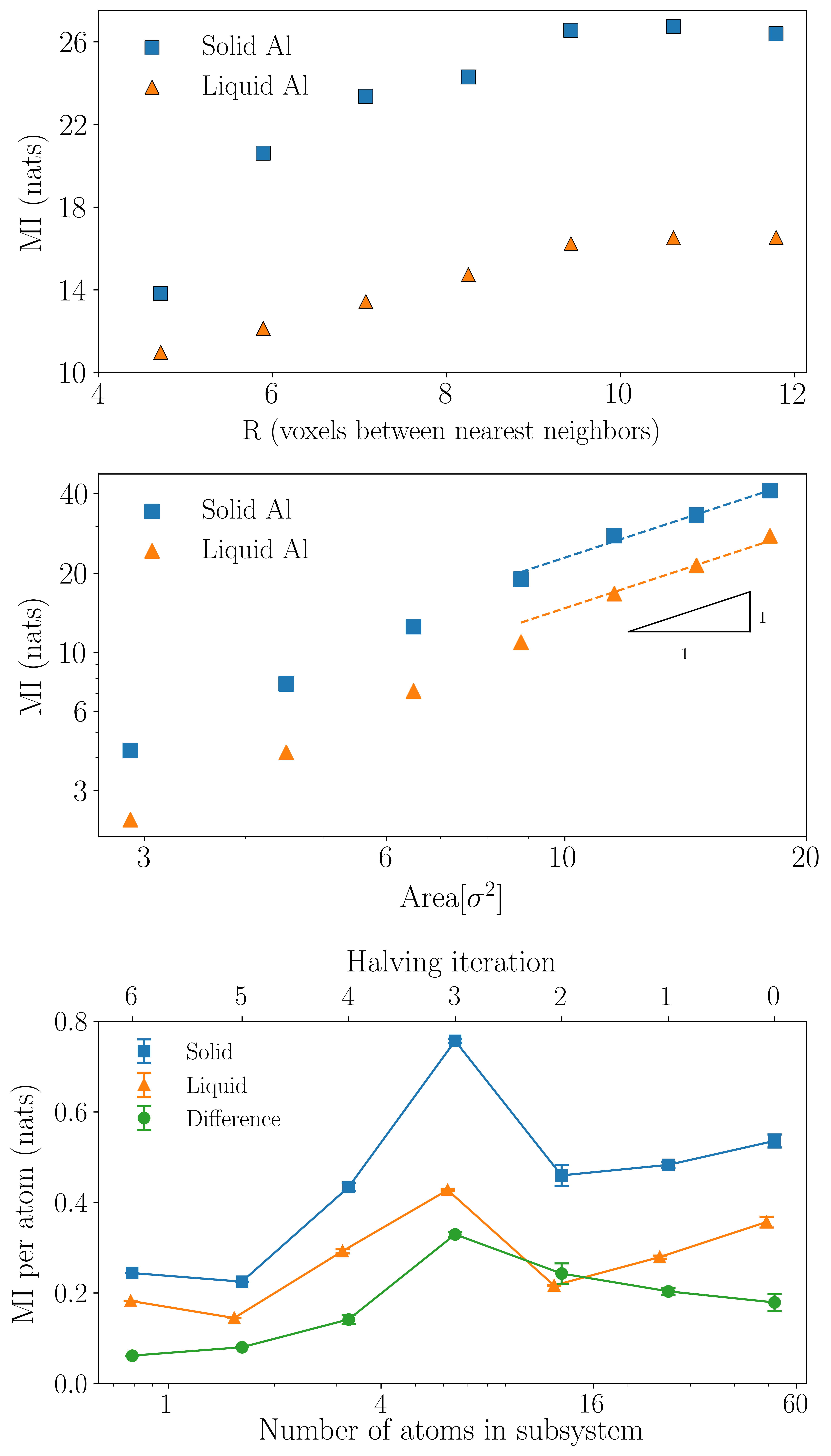}
\caption{\textbf{MI estimations for liquid and crystalline Aluminum} (top)  MI estimation for a system with a fixed physical size as a function of the spatial resolution. The estimate plateaus at high resolutions. (middle) Estimation of MI between two halves of a cubic subsystem, as a function of the interface area. The dashed lines show a linear trendline, showing that large systems obey an area law. 
(bottom) MI density across successive partitions. Starting from the rightmost panel, the system is halved along a different dimension at each step, and MI is computed across the resulting interface.}
\label{fig:res_Al}
\end{figure}

%% file: main.bbl
\begin{thebibliography}{42}%
\makeatletter
\providecommand \@ifxundefined [1]{%
 \@ifx{#1\undefined}
}%
\providecommand \@ifnum [1]{%
 \ifnum #1\expandafter \@firstoftwo
 \else \expandafter \@secondoftwo
 \fi
}%
\providecommand \@ifx [1]{%
 \ifx #1\expandafter \@firstoftwo
 \else \expandafter \@secondoftwo
 \fi
}%
\providecommand \natexlab [1]{#1}%
\providecommand \enquote  [1]{``#1''}%
\providecommand \bibnamefont  [1]{#1}%
\providecommand \bibfnamefont [1]{#1}%
\providecommand \citenamefont [1]{#1}%
\providecommand \href@noop [0]{\@secondoftwo}%
\providecommand \href [0]{\begingroup \@sanitize@url \@href}%
\providecommand \@href[1]{\@@startlink{#1}\@@href}%
\providecommand \@@href[1]{\endgroup#1\@@endlink}%
\providecommand \@sanitize@url [0]{\catcode `\\12\catcode `\$12\catcode `\&12\catcode `\#12\catcode `\^12\catcode `\_12\catcode `\%12\relax}%
\providecommand \@@startlink[1]{}%
\providecommand \@@endlink[0]{}%
\providecommand \url  [0]{\begingroup\@sanitize@url \@url }%
\providecommand \@url [1]{\endgroup\@href {#1}{\urlprefix }}%
\providecommand \urlprefix  [0]{URL }%
\providecommand \Eprint [0]{\href }%
\providecommand \doibase [0]{https://doi.org/}%
\providecommand \selectlanguage [0]{\@gobble}%
\providecommand \bibinfo  [0]{\@secondoftwo}%
\providecommand \bibfield  [0]{\@secondoftwo}%
\providecommand \translation [1]{[#1]}%
\providecommand \BibitemOpen [0]{}%
\providecommand \bibitemStop [0]{}%
\providecommand \bibitemNoStop [0]{.\EOS\space}%
\providecommand \EOS [0]{\spacefactor3000\relax}%
\providecommand \BibitemShut  [1]{\csname bibitem#1\endcsname}%
\let\auto@bib@innerbib\@empty
\bibitem [{\citenamefont {Frenkel}\ and\ \citenamefont {Smit}(2023)}]{frenkel2023understanding}%
  \BibitemOpen
  \bibfield  {author} {\bibinfo {author} {\bibfnamefont {D.}~\bibnamefont {Frenkel}}\ and\ \bibinfo {author} {\bibfnamefont {B.}~\bibnamefont {Smit}},\ }\href@noop {} {\emph {\bibinfo {title} {Understanding molecular simulation: from algorithms to applications}}}\ (\bibinfo  {publisher} {Elsevier},\ \bibinfo {year} {2023})\BibitemShut {NoStop}%
\bibitem [{\citenamefont {Barducci}\ \emph {et~al.}(2008)\citenamefont {Barducci}, \citenamefont {Bussi},\ and\ \citenamefont {Parrinello}}]{barducci2008well}%
  \BibitemOpen
  \bibfield  {author} {\bibinfo {author} {\bibfnamefont {A.}~\bibnamefont {Barducci}}, \bibinfo {author} {\bibfnamefont {G.}~\bibnamefont {Bussi}},\ and\ \bibinfo {author} {\bibfnamefont {M.}~\bibnamefont {Parrinello}},\ }\href@noop {} {\bibfield  {journal} {\bibinfo  {journal} {Physical review letters}\ }\textbf {\bibinfo {volume} {100}},\ \bibinfo {pages} {020603} (\bibinfo {year} {2008})}\BibitemShut {NoStop}%
\bibitem [{\citenamefont {Barducci}\ \emph {et~al.}(2011)\citenamefont {Barducci}, \citenamefont {Bonomi},\ and\ \citenamefont {Parrinello}}]{barducci2011metadynamics}%
  \BibitemOpen
  \bibfield  {author} {\bibinfo {author} {\bibfnamefont {A.}~\bibnamefont {Barducci}}, \bibinfo {author} {\bibfnamefont {M.}~\bibnamefont {Bonomi}},\ and\ \bibinfo {author} {\bibfnamefont {M.}~\bibnamefont {Parrinello}},\ }\href@noop {} {\bibfield  {journal} {\bibinfo  {journal} {Wiley Interdisciplinary Reviews: Computational Molecular Science}\ }\textbf {\bibinfo {volume} {1}},\ \bibinfo {pages} {826} (\bibinfo {year} {2011})}\BibitemShut {NoStop}%
\bibitem [{\citenamefont {Valsson}\ \emph {et~al.}(2016)\citenamefont {Valsson}, \citenamefont {Tiwary},\ and\ \citenamefont {Parrinello}}]{valsson2016enhancing}%
  \BibitemOpen
  \bibfield  {author} {\bibinfo {author} {\bibfnamefont {O.}~\bibnamefont {Valsson}}, \bibinfo {author} {\bibfnamefont {P.}~\bibnamefont {Tiwary}},\ and\ \bibinfo {author} {\bibfnamefont {M.}~\bibnamefont {Parrinello}},\ }\href@noop {} {\bibfield  {journal} {\bibinfo  {journal} {Annual review of physical chemistry}\ }\textbf {\bibinfo {volume} {67}},\ \bibinfo {pages} {159} (\bibinfo {year} {2016})}\BibitemShut {NoStop}%
\bibitem [{\citenamefont {Sutto}\ \emph {et~al.}(2012)\citenamefont {Sutto}, \citenamefont {Marsili},\ and\ \citenamefont {Gervasio}}]{sutto2012new}%
  \BibitemOpen
  \bibfield  {author} {\bibinfo {author} {\bibfnamefont {L.}~\bibnamefont {Sutto}}, \bibinfo {author} {\bibfnamefont {S.}~\bibnamefont {Marsili}},\ and\ \bibinfo {author} {\bibfnamefont {F.~L.}\ \bibnamefont {Gervasio}},\ }\href@noop {} {\bibfield  {journal} {\bibinfo  {journal} {Wiley Interdisciplinary Reviews: Computational Molecular Science}\ }\textbf {\bibinfo {volume} {2}},\ \bibinfo {pages} {771} (\bibinfo {year} {2012})}\BibitemShut {NoStop}%
\bibitem [{\citenamefont {Bussi}\ and\ \citenamefont {Laio}(2020)}]{bussi2020using}%
  \BibitemOpen
  \bibfield  {author} {\bibinfo {author} {\bibfnamefont {G.}~\bibnamefont {Bussi}}\ and\ \bibinfo {author} {\bibfnamefont {A.}~\bibnamefont {Laio}},\ }\href@noop {} {\bibfield  {journal} {\bibinfo  {journal} {Nature Reviews Physics}\ }\textbf {\bibinfo {volume} {2}},\ \bibinfo {pages} {200} (\bibinfo {year} {2020})}\BibitemShut {NoStop}%
\bibitem [{\citenamefont {K{\"a}stner}(2011)}]{kastner2011umbrella}%
  \BibitemOpen
  \bibfield  {author} {\bibinfo {author} {\bibfnamefont {J.}~\bibnamefont {K{\"a}stner}},\ }\href@noop {} {\bibfield  {journal} {\bibinfo  {journal} {Wiley Interdisciplinary Reviews: Computational Molecular Science}\ }\textbf {\bibinfo {volume} {1}},\ \bibinfo {pages} {932} (\bibinfo {year} {2011})}\BibitemShut {NoStop}%
\bibitem [{\citenamefont {Torrie}\ and\ \citenamefont {Valleau}(1977)}]{torrie1977nonphysical}%
  \BibitemOpen
  \bibfield  {author} {\bibinfo {author} {\bibfnamefont {G.~M.}\ \bibnamefont {Torrie}}\ and\ \bibinfo {author} {\bibfnamefont {J.~P.}\ \bibnamefont {Valleau}},\ }\href@noop {} {\bibfield  {journal} {\bibinfo  {journal} {Journal of computational physics}\ }\textbf {\bibinfo {volume} {23}},\ \bibinfo {pages} {187} (\bibinfo {year} {1977})}\BibitemShut {NoStop}%
\bibitem [{\citenamefont {Miao}\ \emph {et~al.}(2015)\citenamefont {Miao}, \citenamefont {Feher},\ and\ \citenamefont {McCammon}}]{miao2015gaussian}%
  \BibitemOpen
  \bibfield  {author} {\bibinfo {author} {\bibfnamefont {Y.}~\bibnamefont {Miao}}, \bibinfo {author} {\bibfnamefont {V.~A.}\ \bibnamefont {Feher}},\ and\ \bibinfo {author} {\bibfnamefont {J.~A.}\ \bibnamefont {McCammon}},\ }\href@noop {} {\bibfield  {journal} {\bibinfo  {journal} {Journal of chemical theory and computation}\ }\textbf {\bibinfo {volume} {11}},\ \bibinfo {pages} {3584} (\bibinfo {year} {2015})}\BibitemShut {NoStop}%
\bibitem [{\citenamefont {Invernizzi}\ and\ \citenamefont {Parrinello}(2020)}]{invernizzi2020rethinking}%
  \BibitemOpen
  \bibfield  {author} {\bibinfo {author} {\bibfnamefont {M.}~\bibnamefont {Invernizzi}}\ and\ \bibinfo {author} {\bibfnamefont {M.}~\bibnamefont {Parrinello}},\ }\href@noop {} {\bibfield  {journal} {\bibinfo  {journal} {The journal of physical chemistry letters}\ }\textbf {\bibinfo {volume} {11}},\ \bibinfo {pages} {2731} (\bibinfo {year} {2020})}\BibitemShut {NoStop}%
\bibitem [{\citenamefont {Invernizzi}\ \emph {et~al.}(2020)\citenamefont {Invernizzi}, \citenamefont {Piaggi},\ and\ \citenamefont {Parrinello}}]{invernizzi2020unified}%
  \BibitemOpen
  \bibfield  {author} {\bibinfo {author} {\bibfnamefont {M.}~\bibnamefont {Invernizzi}}, \bibinfo {author} {\bibfnamefont {P.~M.}\ \bibnamefont {Piaggi}},\ and\ \bibinfo {author} {\bibfnamefont {M.}~\bibnamefont {Parrinello}},\ }\href@noop {} {\bibfield  {journal} {\bibinfo  {journal} {Physical Review X}\ }\textbf {\bibinfo {volume} {10}},\ \bibinfo {pages} {041034} (\bibinfo {year} {2020})}\BibitemShut {NoStop}%
\bibitem [{\citenamefont {Invernizzi}\ and\ \citenamefont {Parrinello}(2022)}]{invernizzi2022exploration}%
  \BibitemOpen
  \bibfield  {author} {\bibinfo {author} {\bibfnamefont {M.}~\bibnamefont {Invernizzi}}\ and\ \bibinfo {author} {\bibfnamefont {M.}~\bibnamefont {Parrinello}},\ }\href@noop {} {\bibfield  {journal} {\bibinfo  {journal} {Journal of Chemical Theory and Computation}\ }\textbf {\bibinfo {volume} {18}},\ \bibinfo {pages} {3988} (\bibinfo {year} {2022})}\BibitemShut {NoStop}%
\bibitem [{\citenamefont {Invernizzi}(2021)}]{Invernizzi2021}%
  \BibitemOpen
  \bibfield  {author} {\bibinfo {author} {\bibfnamefont {M.}~\bibnamefont {Invernizzi}},\ }\bibfield  {journal} {\bibinfo  {journal} {Nuovo Cimento della Societa Italiana di Fisica C}\ }\textbf {\bibinfo {volume} {44}},\ \href {https://doi.org/10.1393/NCC/I2021-21112-8} {10.1393/NCC/I2021-21112-8} (\bibinfo {year} {2021}),\ \Eprint {https://arxiv.org/abs/2101.06991} {arXiv:2101.06991} \BibitemShut {NoStop}%
\bibitem [{\citenamefont {Rogal}\ \emph {et~al.}(2019)\citenamefont {Rogal}, \citenamefont {Schneider},\ and\ \citenamefont {Tuckerman}}]{rogal2019neural}%
  \BibitemOpen
  \bibfield  {author} {\bibinfo {author} {\bibfnamefont {J.}~\bibnamefont {Rogal}}, \bibinfo {author} {\bibfnamefont {E.}~\bibnamefont {Schneider}},\ and\ \bibinfo {author} {\bibfnamefont {M.~E.}\ \bibnamefont {Tuckerman}},\ }\href@noop {} {\bibfield  {journal} {\bibinfo  {journal} {Physical Review Letters}\ }\textbf {\bibinfo {volume} {123}},\ \bibinfo {pages} {245701} (\bibinfo {year} {2019})}\BibitemShut {NoStop}%
\bibitem [{\citenamefont {Callen}(1980)}]{callen1980thermodynamics}%
  \BibitemOpen
  \bibfield  {author} {\bibinfo {author} {\bibfnamefont {H.~B.}\ \bibnamefont {Callen}},\ }\href@noop {} {\bibfield  {journal} {\bibinfo  {journal} {John Wiley\& Sons}\ }\textbf {\bibinfo {volume} {2}} (\bibinfo {year} {1980})}\BibitemShut {NoStop}%
\bibitem [{\citenamefont {Avinery}\ \emph {et~al.}(2019)\citenamefont {Avinery}, \citenamefont {Kornreich},\ and\ \citenamefont {Beck}}]{avinery2019universal}%
  \BibitemOpen
  \bibfield  {author} {\bibinfo {author} {\bibfnamefont {R.}~\bibnamefont {Avinery}}, \bibinfo {author} {\bibfnamefont {M.}~\bibnamefont {Kornreich}},\ and\ \bibinfo {author} {\bibfnamefont {R.}~\bibnamefont {Beck}},\ }\href@noop {} {\bibfield  {journal} {\bibinfo  {journal} {Physical review letters}\ }\textbf {\bibinfo {volume} {123}},\ \bibinfo {pages} {178102} (\bibinfo {year} {2019})}\BibitemShut {NoStop}%
\bibitem [{\citenamefont {Martiniani}\ \emph {et~al.}(2019)\citenamefont {Martiniani}, \citenamefont {Chaikin},\ and\ \citenamefont {Levine}}]{martiniani2019quantifying}%
  \BibitemOpen
  \bibfield  {author} {\bibinfo {author} {\bibfnamefont {S.}~\bibnamefont {Martiniani}}, \bibinfo {author} {\bibfnamefont {P.~M.}\ \bibnamefont {Chaikin}},\ and\ \bibinfo {author} {\bibfnamefont {D.}~\bibnamefont {Levine}},\ }\href@noop {} {\bibfield  {journal} {\bibinfo  {journal} {Physical Review X}\ }\textbf {\bibinfo {volume} {9}},\ \bibinfo {pages} {011031} (\bibinfo {year} {2019})}\BibitemShut {NoStop}%
\bibitem [{\citenamefont {Liu}\ and\ \citenamefont {Simine}(2024)}]{liu2024deltagzip}%
  \BibitemOpen
  \bibfield  {author} {\bibinfo {author} {\bibfnamefont {T.}~\bibnamefont {Liu}}\ and\ \bibinfo {author} {\bibfnamefont {L.}~\bibnamefont {Simine}},\ }\href@noop {} {\bibfield  {journal} {\bibinfo  {journal} {Journal of Chemical Information and Modeling}\ }\textbf {\bibinfo {volume} {64}},\ \bibinfo {pages} {5617} (\bibinfo {year} {2024})}\BibitemShut {NoStop}%
\bibitem [{\citenamefont {Zu}\ \emph {et~al.}(2020)\citenamefont {Zu}, \citenamefont {Bupathy}, \citenamefont {Frenkel},\ and\ \citenamefont {Sastry}}]{Zu2020}%
  \BibitemOpen
  \bibfield  {author} {\bibinfo {author} {\bibfnamefont {M.}~\bibnamefont {Zu}}, \bibinfo {author} {\bibfnamefont {A.}~\bibnamefont {Bupathy}}, \bibinfo {author} {\bibfnamefont {D.}~\bibnamefont {Frenkel}},\ and\ \bibinfo {author} {\bibfnamefont {S.}~\bibnamefont {Sastry}},\ }\href {https://doi.org/10.1088/1742-5468/ab684b} {\bibfield  {journal} {\bibinfo  {journal} {Journal of Statistical Mechanics: Theory and Experiment}\ }\textbf {\bibinfo {volume} {2020}},\ \bibinfo {pages} {023204} (\bibinfo {year} {2020})}\BibitemShut {NoStop}%
\bibitem [{\citenamefont {Sorkin}\ \emph {et~al.}(2023{\natexlab{a}})\citenamefont {Sorkin}, \citenamefont {Be’er}, \citenamefont {Diamant},\ and\ \citenamefont {Ariel}}]{sorkin2023detecting}%
  \BibitemOpen
  \bibfield  {author} {\bibinfo {author} {\bibfnamefont {B.}~\bibnamefont {Sorkin}}, \bibinfo {author} {\bibfnamefont {A.}~\bibnamefont {Be’er}}, \bibinfo {author} {\bibfnamefont {H.}~\bibnamefont {Diamant}},\ and\ \bibinfo {author} {\bibfnamefont {G.}~\bibnamefont {Ariel}},\ }\href@noop {} {\bibfield  {journal} {\bibinfo  {journal} {Soft Matter}\ }\textbf {\bibinfo {volume} {19}},\ \bibinfo {pages} {5118} (\bibinfo {year} {2023}{\natexlab{a}})}\BibitemShut {NoStop}%
\bibitem [{\citenamefont {Sorkin}\ \emph {et~al.}(2023{\natexlab{b}})\citenamefont {Sorkin}, \citenamefont {Ricouvier}, \citenamefont {Diamant},\ and\ \citenamefont {Ariel}}]{sorkin2023resolving}%
  \BibitemOpen
  \bibfield  {author} {\bibinfo {author} {\bibfnamefont {B.}~\bibnamefont {Sorkin}}, \bibinfo {author} {\bibfnamefont {J.}~\bibnamefont {Ricouvier}}, \bibinfo {author} {\bibfnamefont {H.}~\bibnamefont {Diamant}},\ and\ \bibinfo {author} {\bibfnamefont {G.}~\bibnamefont {Ariel}},\ }\href@noop {} {\bibfield  {journal} {\bibinfo  {journal} {Physical Review E}\ }\textbf {\bibinfo {volume} {107}},\ \bibinfo {pages} {014138} (\bibinfo {year} {2023}{\natexlab{b}})}\BibitemShut {NoStop}%
\bibitem [{\citenamefont {Gelman}\ and\ \citenamefont {Cohen}(2024)}]{gelman2024nonequilibrium}%
  \BibitemOpen
  \bibfield  {author} {\bibinfo {author} {\bibfnamefont {S.~D.}\ \bibnamefont {Gelman}}\ and\ \bibinfo {author} {\bibfnamefont {G.}~\bibnamefont {Cohen}},\ }\href@noop {} {\bibfield  {journal} {\bibinfo  {journal} {arXiv preprint arXiv:2405.04877}\ } (\bibinfo {year} {2024})}\BibitemShut {NoStop}%
\bibitem [{\citenamefont {No{\'e}}\ \emph {et~al.}(2019)\citenamefont {No{\'e}}, \citenamefont {Olsson}, \citenamefont {K{\"o}hler},\ and\ \citenamefont {Wu}}]{noe2019boltzmann}%
  \BibitemOpen
  \bibfield  {author} {\bibinfo {author} {\bibfnamefont {F.}~\bibnamefont {No{\'e}}}, \bibinfo {author} {\bibfnamefont {S.}~\bibnamefont {Olsson}}, \bibinfo {author} {\bibfnamefont {J.}~\bibnamefont {K{\"o}hler}},\ and\ \bibinfo {author} {\bibfnamefont {H.}~\bibnamefont {Wu}},\ }\href@noop {} {\bibfield  {journal} {\bibinfo  {journal} {Science}\ }\textbf {\bibinfo {volume} {365}},\ \bibinfo {pages} {eaaw1147} (\bibinfo {year} {2019})}\BibitemShut {NoStop}%
\bibitem [{\citenamefont {Petersen}\ \emph {et~al.}(2023)\citenamefont {Petersen}, \citenamefont {Roig},\ and\ \citenamefont {Covino}}]{petersen2023dynamicsdiffusion}%
  \BibitemOpen
  \bibfield  {author} {\bibinfo {author} {\bibfnamefont {M.}~\bibnamefont {Petersen}}, \bibinfo {author} {\bibfnamefont {G.}~\bibnamefont {Roig}},\ and\ \bibinfo {author} {\bibfnamefont {R.}~\bibnamefont {Covino}},\ }in\ \href {https://openreview.net/forum?id=pwYCCq4xAf} {\emph {\bibinfo {booktitle} {NeurIPS 2023 AI for Science Workshop}}}\ (\bibinfo {year} {2023})\BibitemShut {NoStop}%
\bibitem [{\citenamefont {Klein}\ and\ \citenamefont {No{\'e}}(2024)}]{klein2024transferable}%
  \BibitemOpen
  \bibfield  {author} {\bibinfo {author} {\bibfnamefont {L.}~\bibnamefont {Klein}}\ and\ \bibinfo {author} {\bibfnamefont {F.}~\bibnamefont {No{\'e}}},\ }\href@noop {} {\bibfield  {journal} {\bibinfo  {journal} {arXiv preprint arXiv:2406.14426}\ } (\bibinfo {year} {2024})}\BibitemShut {NoStop}%
\bibitem [{\citenamefont {Nir}\ \emph {et~al.}(2020)\citenamefont {Nir}, \citenamefont {Sela}, \citenamefont {Beck},\ and\ \citenamefont {Bar-Sinai}}]{Nir_2020}%
  \BibitemOpen
  \bibfield  {author} {\bibinfo {author} {\bibfnamefont {A.}~\bibnamefont {Nir}}, \bibinfo {author} {\bibfnamefont {E.}~\bibnamefont {Sela}}, \bibinfo {author} {\bibfnamefont {R.}~\bibnamefont {Beck}},\ and\ \bibinfo {author} {\bibfnamefont {Y.}~\bibnamefont {Bar-Sinai}},\ }\href {https://doi.org/10.1073/pnas.2017042117} {\bibfield  {journal} {\bibinfo  {journal} {Proceedings of the National Academy of Sciences}\ }\textbf {\bibinfo {volume} {117}},\ \bibinfo {pages} {30234–30240} (\bibinfo {year} {2020})}\BibitemShut {NoStop}%
\bibitem [{\citenamefont {Belghazi}\ \emph {et~al.}(2018)\citenamefont {Belghazi}, \citenamefont {Baratin}, \citenamefont {Rajeshwar}, \citenamefont {Ozair}, \citenamefont {Bengio}, \citenamefont {Courville},\ and\ \citenamefont {Hjelm}}]{belghazi2021minemutualinformationneural}%
  \BibitemOpen
  \bibfield  {author} {\bibinfo {author} {\bibfnamefont {M.~I.}\ \bibnamefont {Belghazi}}, \bibinfo {author} {\bibfnamefont {A.}~\bibnamefont {Baratin}}, \bibinfo {author} {\bibfnamefont {S.}~\bibnamefont {Rajeshwar}}, \bibinfo {author} {\bibfnamefont {S.}~\bibnamefont {Ozair}}, \bibinfo {author} {\bibfnamefont {Y.}~\bibnamefont {Bengio}}, \bibinfo {author} {\bibfnamefont {A.}~\bibnamefont {Courville}},\ and\ \bibinfo {author} {\bibfnamefont {D.}~\bibnamefont {Hjelm}},\ }in\ \href@noop {} {\emph {\bibinfo {booktitle} {International conference on machine learning}}}\ (\bibinfo {organization} {PMLR},\ \bibinfo {year} {2018})\ pp.\ \bibinfo {pages} {531--540}\BibitemShut {NoStop}%
\bibitem [{\citenamefont {Donsker}\ and\ \citenamefont {Varadhan}(1975)}]{donsker1975asymptotic}%
  \BibitemOpen
  \bibfield  {author} {\bibinfo {author} {\bibfnamefont {M.~D.}\ \bibnamefont {Donsker}}\ and\ \bibinfo {author} {\bibfnamefont {S.~S.}\ \bibnamefont {Varadhan}},\ }\href@noop {} {\bibfield  {journal} {\bibinfo  {journal} {Communications on pure and applied mathematics}\ }\textbf {\bibinfo {volume} {28}},\ \bibinfo {pages} {1} (\bibinfo {year} {1975})}\BibitemShut {NoStop}%
\bibitem [{\citenamefont {Choi}\ and\ \citenamefont {Lee}(2022)}]{choi2022combating}%
  \BibitemOpen
  \bibfield  {author} {\bibinfo {author} {\bibfnamefont {K.}~\bibnamefont {Choi}}\ and\ \bibinfo {author} {\bibfnamefont {S.}~\bibnamefont {Lee}},\ }in\ \href@noop {} {\emph {\bibinfo {booktitle} {Uncertainty in Artificial Intelligence}}}\ (\bibinfo {organization} {PMLR},\ \bibinfo {year} {2022})\ pp.\ \bibinfo {pages} {411--421}\BibitemShut {NoStop}%
\bibitem [{\citenamefont {Wolf}\ \emph {et~al.}(2008)\citenamefont {Wolf}, \citenamefont {Verstraete}, \citenamefont {Hastings},\ and\ \citenamefont {Cirac}}]{wolf2008area}%
  \BibitemOpen
  \bibfield  {author} {\bibinfo {author} {\bibfnamefont {M.~M.}\ \bibnamefont {Wolf}}, \bibinfo {author} {\bibfnamefont {F.}~\bibnamefont {Verstraete}}, \bibinfo {author} {\bibfnamefont {M.~B.}\ \bibnamefont {Hastings}},\ and\ \bibinfo {author} {\bibfnamefont {J.~I.}\ \bibnamefont {Cirac}},\ }\href@noop {} {\bibfield  {journal} {\bibinfo  {journal} {Physical review letters}\ }\textbf {\bibinfo {volume} {100}},\ \bibinfo {pages} {070502} (\bibinfo {year} {2008})}\BibitemShut {NoStop}%
\bibitem [{\citenamefont {Piaggi}\ \emph {et~al.}(2017)\citenamefont {Piaggi}, \citenamefont {Valsson},\ and\ \citenamefont {Parrinello}}]{piaggi2017enhancing}%
  \BibitemOpen
  \bibfield  {author} {\bibinfo {author} {\bibfnamefont {P.~M.}\ \bibnamefont {Piaggi}}, \bibinfo {author} {\bibfnamefont {O.}~\bibnamefont {Valsson}},\ and\ \bibinfo {author} {\bibfnamefont {M.}~\bibnamefont {Parrinello}},\ }\href@noop {} {\bibfield  {journal} {\bibinfo  {journal} {Physical review letters}\ }\textbf {\bibinfo {volume} {119}},\ \bibinfo {pages} {015701} (\bibinfo {year} {2017})}\BibitemShut {NoStop}%
\bibitem [{\citenamefont {Thompson}\ \emph {et~al.}(2022)\citenamefont {Thompson}, \citenamefont {Aktulga}, \citenamefont {Berger}, \citenamefont {Bolintineanu}, \citenamefont {Brown}, \citenamefont {Crozier}, \citenamefont {In't~Veld}, \citenamefont {Kohlmeyer}, \citenamefont {Moore}, \citenamefont {Nguyen} \emph {et~al.}}]{thompson2022lammps}%
  \BibitemOpen
  \bibfield  {author} {\bibinfo {author} {\bibfnamefont {A.~P.}\ \bibnamefont {Thompson}}, \bibinfo {author} {\bibfnamefont {H.~M.}\ \bibnamefont {Aktulga}}, \bibinfo {author} {\bibfnamefont {R.}~\bibnamefont {Berger}}, \bibinfo {author} {\bibfnamefont {D.~S.}\ \bibnamefont {Bolintineanu}}, \bibinfo {author} {\bibfnamefont {W.~M.}\ \bibnamefont {Brown}}, \bibinfo {author} {\bibfnamefont {P.~S.}\ \bibnamefont {Crozier}}, \bibinfo {author} {\bibfnamefont {P.~J.}\ \bibnamefont {In't~Veld}}, \bibinfo {author} {\bibfnamefont {A.}~\bibnamefont {Kohlmeyer}}, \bibinfo {author} {\bibfnamefont {S.~G.}\ \bibnamefont {Moore}}, \bibinfo {author} {\bibfnamefont {T.~D.}\ \bibnamefont {Nguyen}}, \emph {et~al.},\ }\href@noop {} {\bibfield  {journal} {\bibinfo  {journal} {Computer physics communications}\ }\textbf {\bibinfo {volume} {271}},\ \bibinfo {pages} {108171} (\bibinfo {year} {2022})}\BibitemShut {NoStop}%
\bibitem [{\citenamefont {Bonomi}\ \emph {et~al.}(2009)\citenamefont {Bonomi}, \citenamefont {Branduardi}, \citenamefont {Bussi}, \citenamefont {Camilloni}, \citenamefont {Provasi}, \citenamefont {Raiteri}, \citenamefont {Donadio}, \citenamefont {Marinelli}, \citenamefont {Pietrucci}, \citenamefont {Broglia} \emph {et~al.}}]{bonomi2009plumed}%
  \BibitemOpen
  \bibfield  {author} {\bibinfo {author} {\bibfnamefont {M.}~\bibnamefont {Bonomi}}, \bibinfo {author} {\bibfnamefont {D.}~\bibnamefont {Branduardi}}, \bibinfo {author} {\bibfnamefont {G.}~\bibnamefont {Bussi}}, \bibinfo {author} {\bibfnamefont {C.}~\bibnamefont {Camilloni}}, \bibinfo {author} {\bibfnamefont {D.}~\bibnamefont {Provasi}}, \bibinfo {author} {\bibfnamefont {P.}~\bibnamefont {Raiteri}}, \bibinfo {author} {\bibfnamefont {D.}~\bibnamefont {Donadio}}, \bibinfo {author} {\bibfnamefont {F.}~\bibnamefont {Marinelli}}, \bibinfo {author} {\bibfnamefont {F.}~\bibnamefont {Pietrucci}}, \bibinfo {author} {\bibfnamefont {R.~A.}\ \bibnamefont {Broglia}}, \emph {et~al.},\ }\href@noop {} {\bibfield  {journal} {\bibinfo  {journal} {Computer Physics Communications}\ }\textbf {\bibinfo {volume} {180}},\ \bibinfo {pages} {1961} (\bibinfo {year} {2009})}\BibitemShut {NoStop}%
\bibitem [{plu(2019)}]{plumed2019promoting}%
  \BibitemOpen
  \href@noop {} {\bibfield  {journal} {\bibinfo  {journal} {Nature methods}\ }\textbf {\bibinfo {volume} {16}},\ \bibinfo {pages} {670} (\bibinfo {year} {2019})}\BibitemShut {NoStop}%
\bibitem [{\citenamefont {Tribello}\ \emph {et~al.}(2014)\citenamefont {Tribello}, \citenamefont {Bonomi}, \citenamefont {Branduardi}, \citenamefont {Camilloni},\ and\ \citenamefont {Bussi}}]{tribello2014plumed}%
  \BibitemOpen
  \bibfield  {author} {\bibinfo {author} {\bibfnamefont {G.~A.}\ \bibnamefont {Tribello}}, \bibinfo {author} {\bibfnamefont {M.}~\bibnamefont {Bonomi}}, \bibinfo {author} {\bibfnamefont {D.}~\bibnamefont {Branduardi}}, \bibinfo {author} {\bibfnamefont {C.}~\bibnamefont {Camilloni}},\ and\ \bibinfo {author} {\bibfnamefont {G.}~\bibnamefont {Bussi}},\ }\href@noop {} {\bibfield  {journal} {\bibinfo  {journal} {Computer physics communications}\ }\textbf {\bibinfo {volume} {185}},\ \bibinfo {pages} {604} (\bibinfo {year} {2014})}\BibitemShut {NoStop}%
\bibitem [{\citenamefont {Bussi}\ \emph {et~al.}(2007)\citenamefont {Bussi}, \citenamefont {Donadio},\ and\ \citenamefont {Parrinello}}]{bussi2007canonical}%
  \BibitemOpen
  \bibfield  {author} {\bibinfo {author} {\bibfnamefont {G.}~\bibnamefont {Bussi}}, \bibinfo {author} {\bibfnamefont {D.}~\bibnamefont {Donadio}},\ and\ \bibinfo {author} {\bibfnamefont {M.}~\bibnamefont {Parrinello}},\ }\href@noop {} {\bibfield  {journal} {\bibinfo  {journal} {The Journal of chemical physics}\ }\textbf {\bibinfo {volume} {126}} (\bibinfo {year} {2007})}\BibitemShut {NoStop}%
\bibitem [{\citenamefont {Parrinello}\ and\ \citenamefont {Rahman}(1981)}]{parrinello1981polymorphic}%
  \BibitemOpen
  \bibfield  {author} {\bibinfo {author} {\bibfnamefont {M.}~\bibnamefont {Parrinello}}\ and\ \bibinfo {author} {\bibfnamefont {A.}~\bibnamefont {Rahman}},\ }\href@noop {} {\bibfield  {journal} {\bibinfo  {journal} {Journal of Applied physics}\ }\textbf {\bibinfo {volume} {52}},\ \bibinfo {pages} {7182} (\bibinfo {year} {1981})}\BibitemShut {NoStop}%
\bibitem [{\citenamefont {Srivastava}\ \emph {et~al.}(2014)\citenamefont {Srivastava}, \citenamefont {Hinton}, \citenamefont {Krizhevsky}, \citenamefont {Sutskever},\ and\ \citenamefont {Salakhutdinov}}]{srivastava14Dropout}%
  \BibitemOpen
  \bibfield  {author} {\bibinfo {author} {\bibfnamefont {N.}~\bibnamefont {Srivastava}}, \bibinfo {author} {\bibfnamefont {G.}~\bibnamefont {Hinton}}, \bibinfo {author} {\bibfnamefont {A.}~\bibnamefont {Krizhevsky}}, \bibinfo {author} {\bibfnamefont {I.}~\bibnamefont {Sutskever}},\ and\ \bibinfo {author} {\bibfnamefont {R.}~\bibnamefont {Salakhutdinov}},\ }\href {http://jmlr.org/papers/v15/srivastava14a.html} {\bibfield  {journal} {\bibinfo  {journal} {Journal of Machine Learning Research}\ }\textbf {\bibinfo {volume} {15}},\ \bibinfo {pages} {1929} (\bibinfo {year} {2014})}\BibitemShut {NoStop}%
\bibitem [{\citenamefont {Glorot}\ and\ \citenamefont {Bengio}(2010)}]{glorot2010understanding}%
  \BibitemOpen
  \bibfield  {author} {\bibinfo {author} {\bibfnamefont {X.}~\bibnamefont {Glorot}}\ and\ \bibinfo {author} {\bibfnamefont {Y.}~\bibnamefont {Bengio}},\ }in\ \href@noop {} {\emph {\bibinfo {booktitle} {Proceedings of the thirteenth international conference on artificial intelligence and statistics}}}\ (\bibinfo {organization} {JMLR Workshop and Conference Proceedings},\ \bibinfo {year} {2010})\ pp.\ \bibinfo {pages} {249--256}\BibitemShut {NoStop}%
\bibitem [{\citenamefont {Batzner}\ \emph {et~al.}(2022)\citenamefont {Batzner}, \citenamefont {Musaelian}, \citenamefont {Sun}, \citenamefont {Geiger}, \citenamefont {Mailoa}, \citenamefont {Kornbluth}, \citenamefont {Molinari}, \citenamefont {Smidt},\ and\ \citenamefont {Kozinsky}}]{batzner20223}%
  \BibitemOpen
  \bibfield  {author} {\bibinfo {author} {\bibfnamefont {S.}~\bibnamefont {Batzner}}, \bibinfo {author} {\bibfnamefont {A.}~\bibnamefont {Musaelian}}, \bibinfo {author} {\bibfnamefont {L.}~\bibnamefont {Sun}}, \bibinfo {author} {\bibfnamefont {M.}~\bibnamefont {Geiger}}, \bibinfo {author} {\bibfnamefont {J.~P.}\ \bibnamefont {Mailoa}}, \bibinfo {author} {\bibfnamefont {M.}~\bibnamefont {Kornbluth}}, \bibinfo {author} {\bibfnamefont {N.}~\bibnamefont {Molinari}}, \bibinfo {author} {\bibfnamefont {T.~E.}\ \bibnamefont {Smidt}},\ and\ \bibinfo {author} {\bibfnamefont {B.}~\bibnamefont {Kozinsky}},\ }\href@noop {} {\bibfield  {journal} {\bibinfo  {journal} {Nature communications}\ }\textbf {\bibinfo {volume} {13}},\ \bibinfo {pages} {2453} (\bibinfo {year} {2022})}\BibitemShut {NoStop}%
\bibitem [{\citenamefont {Satorras}\ \emph {et~al.}(2021)\citenamefont {Satorras}, \citenamefont {Hoogeboom},\ and\ \citenamefont {Welling}}]{satorras2021n}%
  \BibitemOpen
  \bibfield  {author} {\bibinfo {author} {\bibfnamefont {V.~G.}\ \bibnamefont {Satorras}}, \bibinfo {author} {\bibfnamefont {E.}~\bibnamefont {Hoogeboom}},\ and\ \bibinfo {author} {\bibfnamefont {M.}~\bibnamefont {Welling}},\ }in\ \href@noop {} {\emph {\bibinfo {booktitle} {International conference on machine learning}}}\ (\bibinfo {organization} {PMLR},\ \bibinfo {year} {2021})\ pp.\ \bibinfo {pages} {9323--9332}\BibitemShut {NoStop}%
\bibitem [{\citenamefont {Gasteiger}\ \emph {et~al.}(2020)\citenamefont {Gasteiger}, \citenamefont {Gro{\ss}},\ and\ \citenamefont {G{\"u}nnemann}}]{gasteiger2020directional}%
  \BibitemOpen
  \bibfield  {author} {\bibinfo {author} {\bibfnamefont {J.}~\bibnamefont {Gasteiger}}, \bibinfo {author} {\bibfnamefont {J.}~\bibnamefont {Gro{\ss}}},\ and\ \bibinfo {author} {\bibfnamefont {S.}~\bibnamefont {G{\"u}nnemann}},\ }\href@noop {} {\bibfield  {journal} {\bibinfo  {journal} {arXiv preprint arXiv:2003.03123}\ } (\bibinfo {year} {2020})}\BibitemShut {NoStop}%
\end{thebibliography}%
